\documentclass[referee]{cjaa}           

\usepackage{graphicx}                   
\input{epsf.sty}                        
\input{psfig.sty}                       

\def\gtrsim{\mathrel{\hbox{\rlap{\hbox{\lower4pt\hbox{$\sim$}}}\hbox{$>$}}}}
\def\lesssim{\mathrel{\hbox{\rlap{\hbox{\lower4pt\hbox{$\sim$}}}\hbox{$<$}}}}

\begin{document}

  \renewcommand{\thefootnote}{\alph{footnote}}

\title{GRBs, SGRs by UHE leptons showering, blazing and re-brightening by precessing
 Gamma Jets  in-off axis }

   \volnopage{Vol.0 (200x) No.0, 000--000}      
   \setcounter{page}{1}          

   \author{D. Fargion,
      \inst{1,2}\mailto{daniele.fargion@roma1.infn.it}
      \ M. Grossi
      \inst{1}      }
   \offprints{D. Fargion}                   

   \institute{Physics Department, University of Rome  "La Sapienza", Pl.
A. Moro 2, 00185, Rome,  Italy\\
             \email{daniele.fargion@roma1.infn.it}
        \and INFN, University of Rome  "La Sapienza", Pl. A. Moro 2,
00185, Rome,  Italy}

   \date{Received~~2006  month day; accepted~~2006~~month day}




  \abstract{ A list of questions regarding Gamma Ray Bursts (GRBs) and Soft Gamma Repeaters (SGRs) remain unanswered
   within the Fireball
   and Magnetar scenarios. On the other hand we argue  that a persistent,
   thin (less than few $\mu$sr) precessing and spinning
    gamma jet,  with  the same power of the progenitor supernova (SN) may explain these issues.
The jets may have precessing time scales of a few hours but their
lifetime could be as long as thousands of years. The orientation
of the beam respect to the line of sight plays a key role in this
scenario: the farthest GRB events correspond to a very narrow and
on-axis beam, while for the nearest ones the beam is off-axis.
     Relic  neutron stars,  X-ray pulsars,  with their  spinning and precessing
     jets  are the candidate blazing sources of GRBs and  SGRs respectively.
      These 'delayed' gamma jets  (even weeks or months after the SN) would appear without a  bright optical transient (OT),
     contrary to the rarest events where the SN and the GRB occur at the same time, and the OT is the result of the  baryon load pollution.
      We expect that nearby off-axis GRBs would be accompanied by a chain of OT and radio bumps
      and possible re-brightening, as in GRB030329-SN2003.
      Delayed  jets are observable in the local universe  as  X-Ray Flahes (XRFs) or short GRBs
       and at closer distances   as SGRs and anomalous X-ray Pulsars AXRPs.
       In our scenario the gamma jet is originated by  ultra-relativistic electron pairs
       showering via Synchrotron (or Inverse Compton)
         radiation outside of  the dense SN core (or the strong neutron star magnetic field).
      We propose that the escape of the electron pairs from the inner core occurs thanks to a more penetrating carrier,
      the  relativistic  PeV muon pairs, themselves secondaries
      of inner UHE hadron jets.  Such leptons are almost
      'transparent' when they propagate through the SN
       shells of matter and its radiative background, thus they are able to  decay far away in to electron pairs (and later on) in gamma and neutrinos jets. }



   \authorrunning{D. Fargion ,\ M. Grossi }            
   \titlerunning{GRBs and SGRs by high energy leptons showering in blazing $\gamma$ jets}  

   \maketitle

\section{Introduction:  GRB-SGR open questions}

Why GRBs are so spread in their total  energy, (above $6$ orders
of magnitude) and in their peak energy (quantities positvely
correlated following the so-called Amati correlation;
\cite{Amati}, see also  \cite{Fa99})? Does the Amati law imply
more and more new GRB families?
   Why are the harder  and  more variable GRBs  (\cite{Lazzati},\cite{Fa99}) found at higher redshifts contrary to  expected Hubble law?
    Why does the output power of GRB vary in a  range  (see also \cite{Fa99}) of $8-9$ orders of
    magnitudes with the most powerful events residing at the cosmic edges (\cite{Yo2004})?
   Why
has  it  been possible to find in the local universe($40-150 Mpc
$) at least
 two nearby events (GRB980425 at z=0.008 and recent GRB060218 at z=0.03)\footnote{The very recent discovers of a second nearest GRB060218 and of an extremely short GRBs,
    but with a longevous life X-ray afterglow, GRB050724,because its long  multi-rebrightening is testing the persistent jet
 activity and  geometrical blazing views; they have been discovered
  just after the Vulcano presentation in May 2005, but they are introduced here to update the article. }?  Most GRBs should be located at $z \geq 1$,
 (\cite{Fa99}). Why are these two GRBs  so much
under-luminous (\cite{Fa99})? Why are they  so slow? Why do their
afterglows show so many bumps and re-brightening as the well-known
third nearest event, GRB030329?
     Why do not many GRB curves show monotonic  decay (an obvious consequence of a one-shot explosive event),
   rather they often show  sudden re-brightening or bumpy afterglows at
different time scales and wavelengths (\cite{Stanek};
\cite{DaF03}; see e.g. GRB 050502B, \cite{Falcone})  ?
    Why have there been a few GRBs and SGRs whose spectra and time structure
    are almost identical  if their origin is so different  (beamed explosion for GRB versus
     isotropic magnetar) (\cite{Fa99}, \cite{Woo99})?
       How can a jetted fireball (with an opening angle of $5^o - 10^o$) release a
       power  nearly 6 orders of magnitude more energetic than the corresponding isotropic SN?  How can re-brightening take place in the  X-ray  and optical afterglows
       (\cite{DaF03})?   How can some ($\sim$ $6 \%$) of the  GRBs (or a few SGRs)
        survive the 'tiny'  (but still extremely powerful)
        explosion of its  $precursor$
     without any consequences, and then explode, catastrophically, a few
     minutes   later?   In such a  scenario, how could the very recent GRB 060124 (at redshift z=2.3)
     be  preceded by a $10$ minutes precursor, and then being able to produce  multiple bursts hundreds of times brighter?
      Why do not huge SGRs, such as SGR1806-20, show evidence of the loss  of angular velocity,    while their hypothetical magnetic energy reservoir has been largely
       exhausted?
     Why do SGR1806 radio afterglows show a mysterious two-bump radio curve implying additional
     energy injection many days later? In this connection why are the
     GRB021004 light curves (from X to radio) calling for an
     early and late energy injection?
     Why has the SGR180620  polarization curve been changing angle radically in
     short ($\sim$ days) timescale?  Why is the short GRB050724 able to bump  and re-bright a day after the main burst \cite{Campana}?
     Once these major questions are addressed and (in our opinion) mostly solved by our
        precessing gamma jet model, a final  question still remains,
        calling for a radical assumption on the thin precessing
        gamma jet: how can an ultra-relativistic electron beam (in any kind of Jet models) survive the SN background and
     dense matter layers and escape in the outer space while remaining  collimated?

     Such questions are ignored in most Fireball models that try to
     fit the very different GRB afterglow light curves with  polynomial
     ad-hoc curves and-or with  unrealistic shell mass
     redistribution around the GRB event. Their solution forces us
     more and more  to the precessing Gamma Jet model fed by the PeV-TeV lepton
     showering discussed below. As we will show, the thin gamma
      precessing jet is indeed made by a chain of primary
      processes (PeV muon pair bundles decaying into
      electrons and then radiating via synchrotron radiation
      ), requiring an inner  ultra-relativistic jet inside the
      source.



\section{Blazing Precessing jets in GRBs and SGRs }

\begin{figure}
   \vspace{2mm}
   \begin{center}
   \includegraphics[width=2.6in]{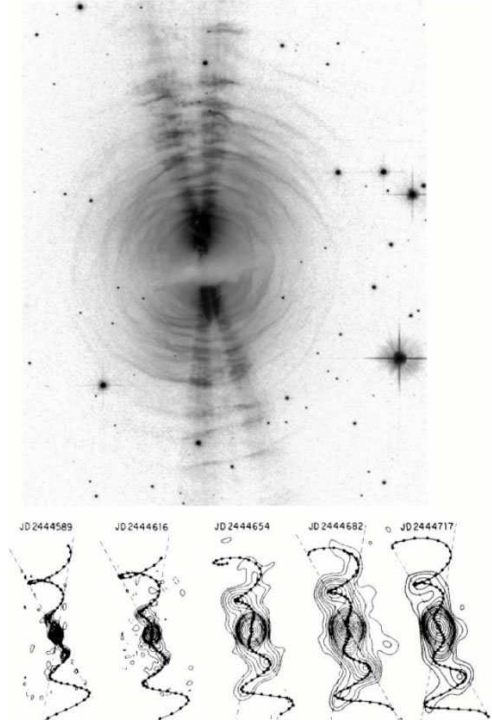}
\caption{{\em Up:} The Egg Nebula, whose shape might be explained
as the conical section of a twin precessing jet interacting with
the surrounding  cloud of ejected gas. {\em Down:} The similar
observed structure of the outflows from the microquasar SS433. A
kinematic model of the time evolution of two oppositely directed
precessing jets is overlaid on the radio contours (from Blundell
\& Bowler 2005).} \label{SS433}
\end{center}
\end{figure}


\begin{figure}
\includegraphics[width=1.3in]{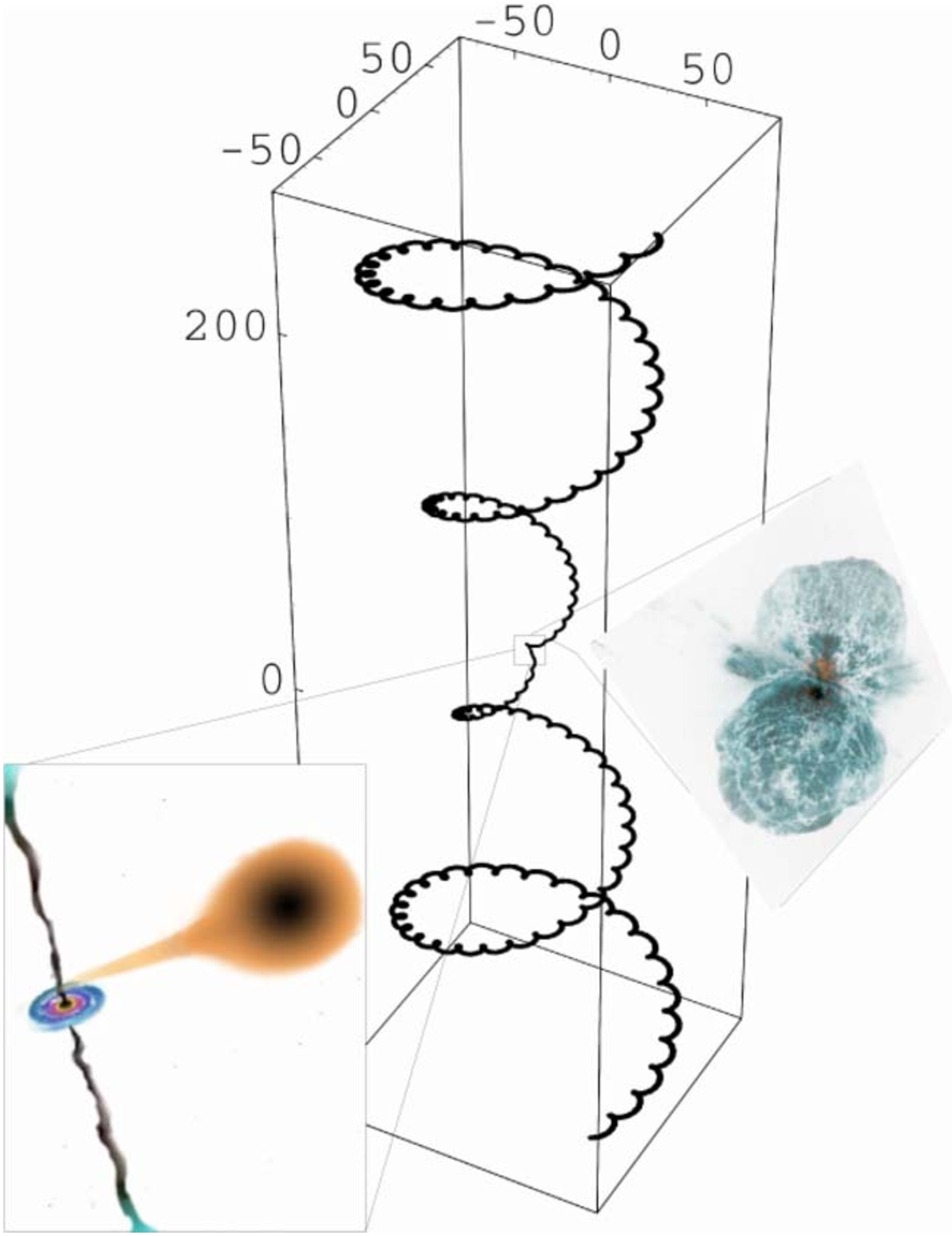}     
\includegraphics[width=1.8in]{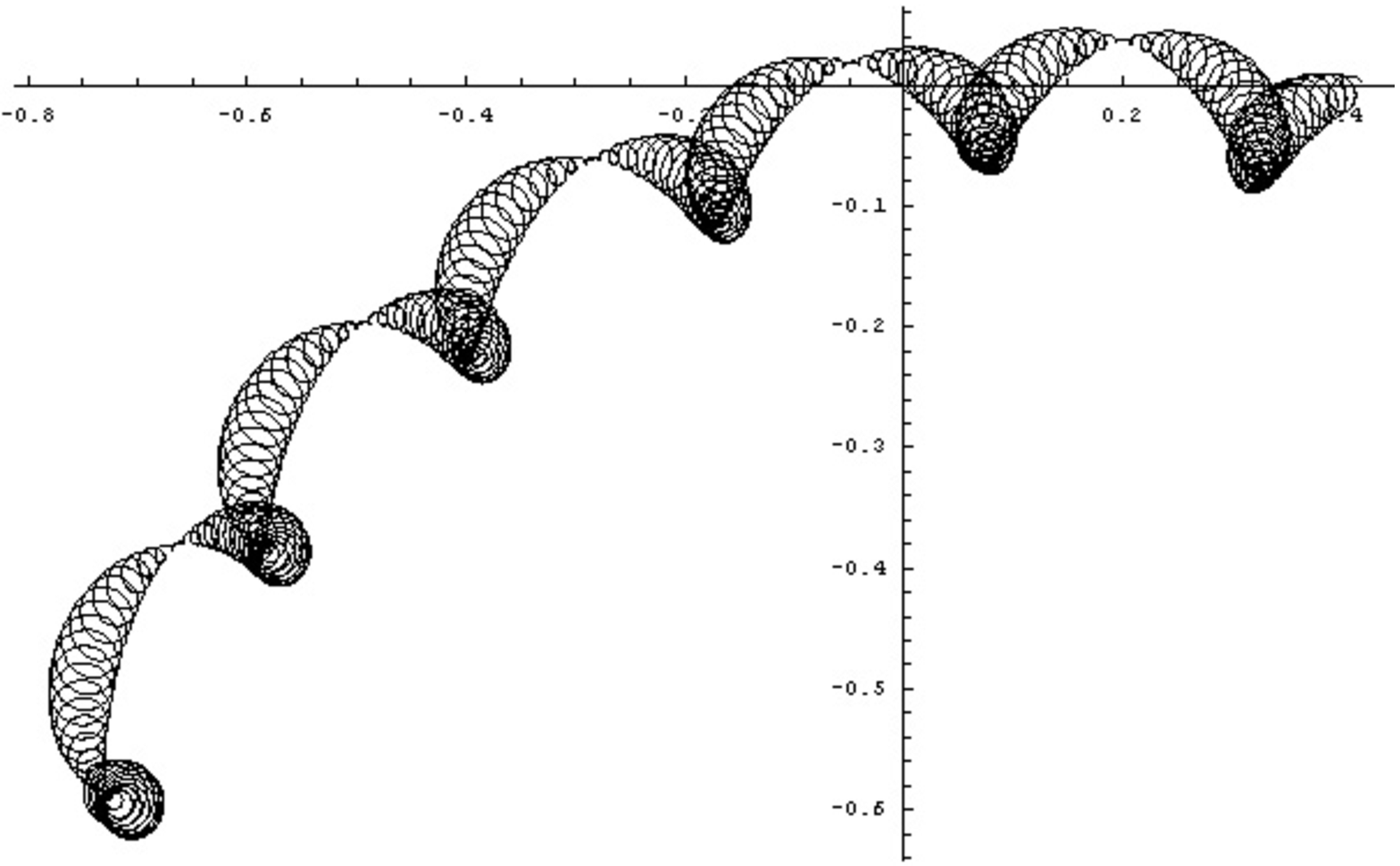}     
\includegraphics[width=1.8in]{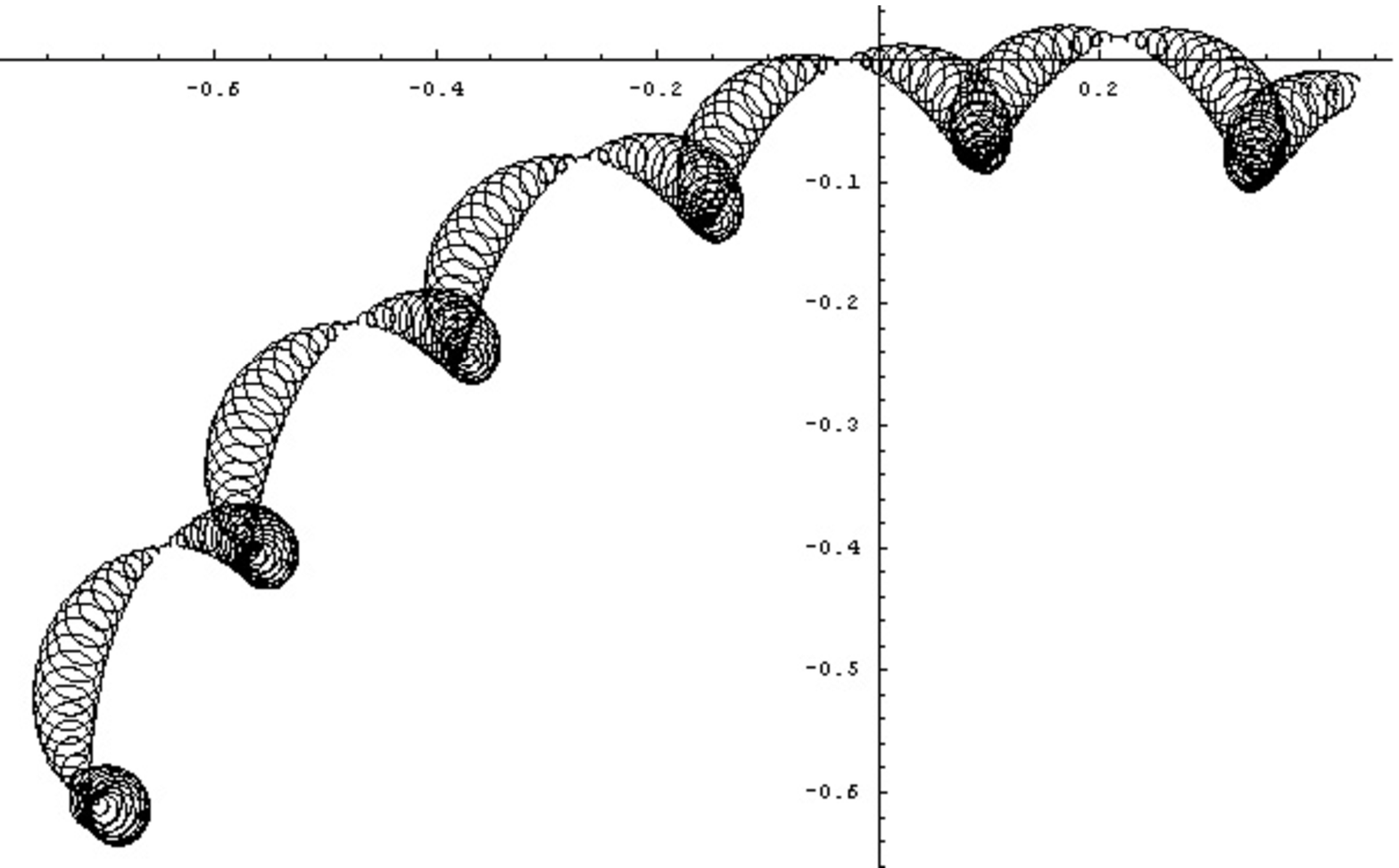}     
\caption{A possible 3D structure view of the precessing jet
obtained with, for instance, a non linear precessing, while
spinning, gamma jet; at its center the "explosive" SN-like event
for a GRB  or a steady binary system for a SGRs where an accretion
disc around a compact object powers a collimated precessing jet.
In the left panel of the figure we show an Herbig Haro - like
object such as HH49, whose spiral jets are describing, at a  lower
energy scale, the ones in micro-quasars such as SS-433. The
Lorentz factor used in the electron pairs jet may reach $\gamma_e
= 10^9$, corresponding to a $\sim PeV$ electron pair energy; its
solid angle nevertheless maybe still asymmetric (small in one side
but not in the other, because magnetic Larmor bending) leading to
a solid angle $\frac{\Delta \Omega}{\Omega} \simeq
10^{-8}-10^{-10}$: two different 2D trajectory  geometry of the
precessing jet while it blazes the observer along the line of
sight: at the center, by a fine tuned beaming to the observer,
while on the right side the slight off-axis flashes of the same
jet; a more powerful but more off-axis event (a few degree away)
would mimic the GRB980425 and GRB060218 soft, long and smooth
X-ray bump } \label{spinning_GRB}
\end{figure}

\begin{figure}
\includegraphics[width=2.5in]{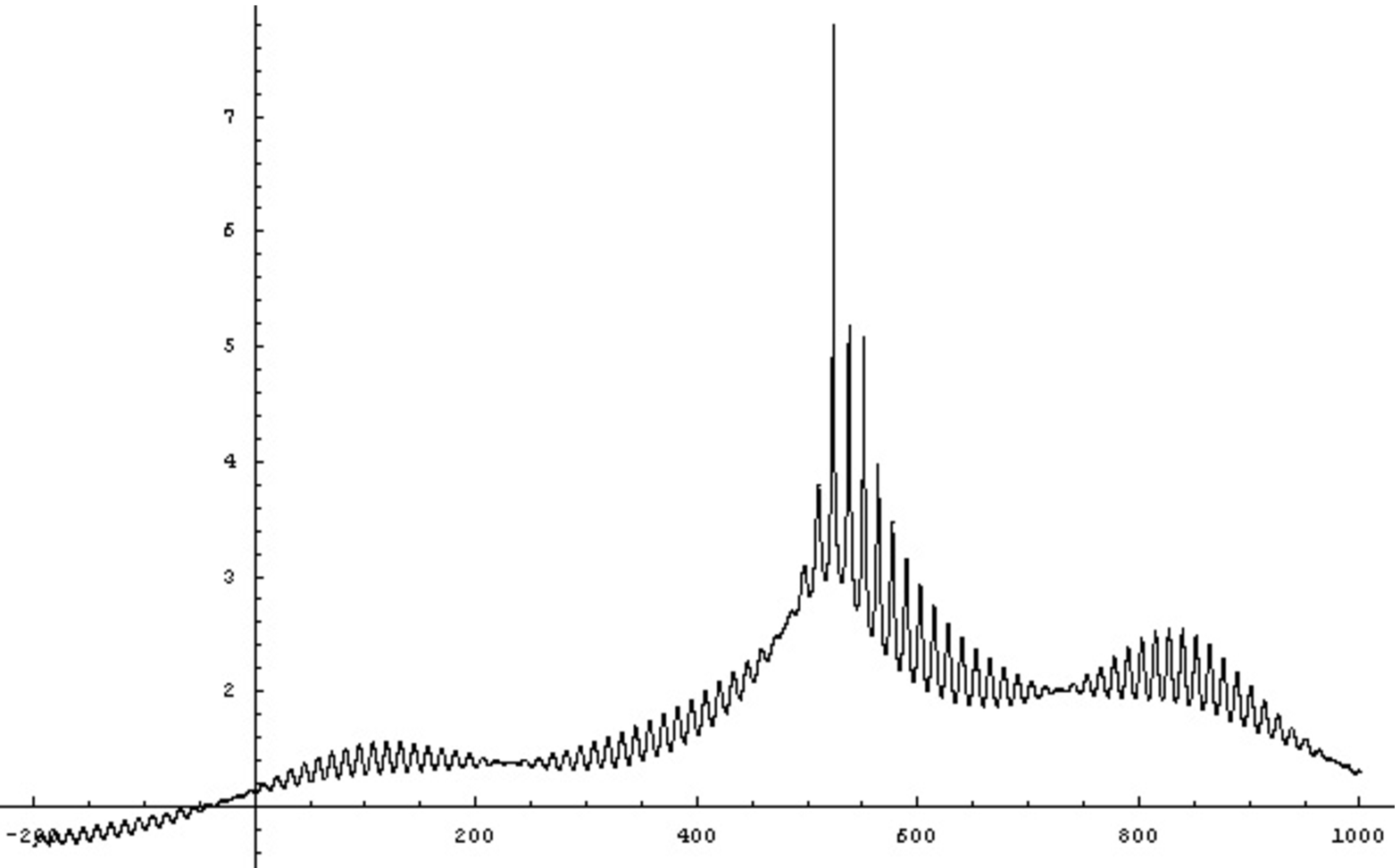}     
\includegraphics[width=2.5in]{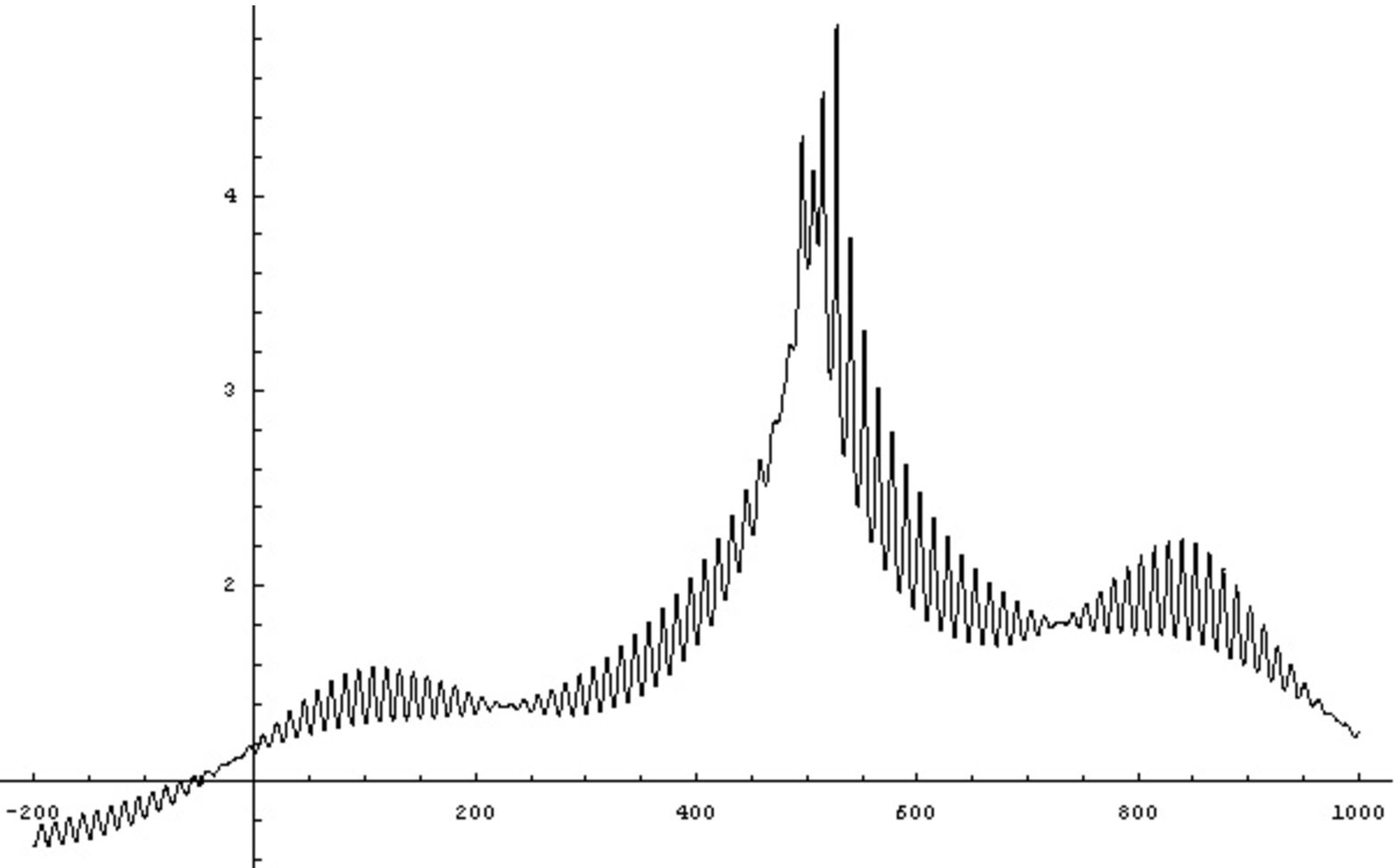}     
\caption{A close up of the two corresponding light curve profile.
Left panel: the multiple oscillatory signals may mimic  the
oscillatory bumps in SGR $\gamma$ and the huge amplification of
giant flare, while the multi-precessing tracks of the jet may lead
to re-brightening and multi-bumps  in the light profile of the GRB
X afterglows as GRB060218. The scale time of the GRB are ruled by
the solid angle of the jet, its impact angle toward the  off-axis
observer, the precessing angular velocities. Right panel: note
that a much off-axis beaming induce a different SGR smoother and
softer profile and a much limited GRB amplification}
\label{profile_SGR}
\end{figure}

The huge GRBs luminosity (up to $10^{54}$ erg s$^{-1}$) may be {
due} to a high collimated, on-axis blazing jet powered by a
Supernova output; the gamma jet is made by relativistic
synchrotron radiation (or ICS) and the inner the jet the harder
and the denser is its output. The harder the photon energy, the
thinner is the jet opening angle $\Delta\theta \simeq
\gamma^{-1}$, $\Delta\Omega \simeq \gamma^{-2}$, $\gamma \simeq
10^4$. The thin solid angle explains the rare SN-GRB connection
and for instance the apparent GRB 990123 extraordinary power
(billions of times the typical SN luminosity). This also explainss
the rarer, because nearer, GRB-SN events such as GRB980425 or
GRB060218, whose jets were off-axis $300 \cdot \gamma^{-1}$, i.e.
a few degrees,(increasing its probability detection by roughly a
hundred thousand times) , but whose GRB luminosity was exactly for
the same reasons extremely low. This beaming selection in larger
volumes explains the puzzling evidence (the Amati correlation) of
harder and apparently powerful GRBs at larger and larger
distances. The statistical selection favors (in wider volumes and
for a wider sample of SN-GRB-jet) the harder and more on-axis
events. A huge unobserved population of far off-axis SN-GRBs are
below the detection thresholds. This Amati correlation remains
unexplained for any isotropic Fireball model and it is in contrast
with the cosmic trend required by the Hubble-Friedmann law: the
further the distances, the larger the redshifts and the softer the
expected
GRB event.
Naturally the farthest events at large redshifts may compensate
duration with time doppler shift. In our opinion to make GRB-SN in
nearly energy equipartition the jet must be very collimated
$\frac{\Omega}{\Delta\Omega}\simeq 10^{8}-10^{10}$ (Fargion,Salis
1995;Fargion 1999; Fargion,Grossi 2005). In order to fit the
statistics between GRB-SN rates, the jet must have a decaying
activity ($ \dot{L}\simeq (\frac{t}{t_o})^{-\alpha}$, $\alpha
\simeq 1$), it must survive not just for the observed GRB
duration, but for a much longer timescale, possibly thousands of
time longer, $t_o \simeq 10^4 s$. The late stages of the GRBs
would appear as SGRs.
Indeed similar criticism (against one shot magnetar model) arises
for the surprising giant flare from SGR 1806-20 that occurred on
2004 December 27: if it has been radiated isotropically ({ as
assumed by} the magnetar model), most of (if not all) the magnetic
energy stored in the neutron star NS should have been consumed at
once. This should have been reflected into sudden angular velocity
loss never observed. On the contrary a thin collimated precessing
jet $\dot{E}_{SGR-jet}\simeq 10^{36}-10^{38} erg s^{-1}$, blazing
on-axis, may be the source of such an apparently (the inverse of
the solid beam angle $\frac{\Omega}{\Delta\Omega}\simeq
10^{8}-10^{9}$) huge bursts $\dot{E}_{SGR-Flare}\simeq
10^{38}\cdot \frac{\Omega}{\Delta\Omega} \simeq 10^{47} erg
s^{-1}$ with a moderate  steady jet output power (X-Pulsar,
SS433). This explains the absence of any variation in the
SGR1806-20 period and its time derivative, contrary to any obvious
correlation with the dipole energy loss law.

 In our model, the temporal evolution of the
angle between the jet direction and the rotational axis {of the
NS} can be expressed as $ \theta_1 (t) = \sqrt{\theta_x^2 +
\theta_y^2}$,
where\\

 $ \theta_x(t) =  \sin (\omega_b t + \phi_{b}) + \theta_{psr}
\cdot \sin( \omega_{psr} t + \phi_{psr})\cdot |(\sin( \omega_N t +
\phi_N))| + \theta_s \cdot \sin (\omega_s t+ \phi_{s}) + \theta_N
\cdot \sin (\omega_N t + \phi_N) + \theta_x(0)$  and \\

 $\theta_y(t) =  \theta_a \cdot \sin \omega_0 t  + \cos ( \omega_b t + \phi_{b})+ \theta_{psr} \cdot \cos (\omega_{psr} t +
\phi_{psr})\cdot |(\sin( \omega_N t + \phi_N))| + \theta_s \cdot
\cos (\omega_s t+ \phi_{s}) + \theta_N \cdot \cos (\omega_N t +
\phi_N)) + \theta_y(0)$ \\

(where $\gamma$ is the Lorentz factor of the relativistic
particles {of the jet}, see Table I and Fig.1 and Fig.2. See also
Fargion 1999 and Fargion 2003).
\begin{table}
\begin{tabular}{lll}
\hline \hline
  $\gamma = 10^9$  & $\theta_a=0.2$ & $\omega_a =1.6 \cdot 10^{-8}$ rad/s\\
  $\theta_b=1$ &  $\theta_{psr}$=1.5 $\cdot 10^7$/$\gamma$ & $\theta_N$=$5 \cdot 10^7$/$\gamma$ \\
$\omega_b$=4.9 $\cdot 10^{-4}$ rad/s &  $\omega_{psr}$=0.83 rad/s
& $\omega_N $=1.38 $\cdot 10^{-2}$ rad/s \\
$\phi_{b}=2\pi - 0.44$ &$\phi_{psr}$=$\pi + \pi/4$ & $\phi_N$=3.5
$\pi/2 + \pi/3$ \\
$\phi_s \sim \phi_{psr}$ & $\theta_s$=1.5 $\cdot 10^6$/$\gamma$ & $\omega_s = 25$ rad/s \\
 \hline \hline
\end{tabular}
\caption{The parameters adopted for the jet model represented in
Fig. \ref{spinning_GRB}} \label{jet_parameters}
\end{table}

The simplest  way to produce the $\gamma$ emission  would be by IC
of GeVs electron pairs onto thermal infra-red photons. Also
electromagnetic showering of PeV electron pairs by synchrotron
emission in galactic fields, ($e^{\pm}$ from muon decay) may be
the progenitor of the $\gamma$ blazing jet. However, the main
difficulty for a jet of GeV electrons is that their propagation
through the SN radiation field is highly suppressed (Fig. 4, left
panel). UHE muons ($E_{\mu} \gtrsim$ PeV) instead are
characterised by a longer interaction length either with the
circum-stellar matter and the radiation field, thus they have the
advantage to avoid the opacity of the star and escape the dense
GRB-SN isotropic radiation field (Fargion,Grossi 2005). 
Here we propose also the emission of SGRs is due to a primary
hadronic jet producing ultra relativistic $e^{\pm}$ (1 - 10 PeV)
from hundreds PeV pions, $\pi\rightarrow \mu \rightarrow e$, (as
well as  EeV neutron decay in flight): primary protons can be
accelerated by the large magnetic field of the NS up to EeV
energy. The protons could emit directly soft gamma rays via
synchrotron radiation with the galactic magnetic field
($E_{\gamma}^p \simeq 10 (E_p/EeV)^2 (B/2.5 \cdot 10^{-6} \, G)$
keV), but the efficiency is poor because of the too long timescale
of proton synchrotron interactions. By interacting with the local
galactic magnetic field relativistic pair electrons lose energy
via synchrotron radiation, $ E_{\gamma}^{sync} \simeq 4.2 \times
10^6 \left(\frac{E_e}{5 \cdot 10^{15} \: eV} \right)^2
\left(\frac{B}{2.5 \cdot 10^{-6} \; G} \right) \: eV$, { with a
characteristic timescale}  $ t^{sync} \simeq 1.3 \times 10^{10}
\left(\frac{E_{e}}{5 \cdot 10^{15} eV} \right)^{-1}
\left(\frac{B}{2.5 \cdot 10^{-6} \, G} \right)^{-2} \: s $.

{ This mechanism  would produce} a few hundreds keV radiation as
it is observed in the intense $\gamma$-ray flare from SGR 1806-20.
The Larmor radius is about two orders of magnitude smaller than
the synchrotron interaction length and this may imply that the
aperture of the jet is spread by the magnetic field, $
\frac{R_L}{c} \simeq 4.1 \times 10^{8}
  \left(\frac{E_{e}}{5 \cdot 10^{15} eV} \right)
\left(\frac{B}{2.5 \cdot 10^{-6} \, G} \right)^{-1} \: s$. In
particular a thin ($\Delta \Omega \simeq 10^{-9}-10^{-10} $ $sr$)
precessing jet  from a pulsar may naturally explain the negligible
variation of the spin frequency $\nu=1/P$ after the giant flare
($\Delta \nu < 10^{-5}$ Hz). Indeed it seems quite unlucky that a
huge ($E_{Flare} \simeq 5 \cdot 10^{46} erg$) explosive event (as
the needed mini-fireball by a magnetar model; Duncan et al. 1992)
is not leaving any trace in the rotational energy of the SGR
1806-20, $ E_{rot}= \frac{1}{2} I_{NS} \omega^2 \simeq 3.6 \cdot
10^{44} \frac{P}{7.5 s}^{-2} \left( \frac{I_{NS}}{10^{45} g \,
cm^2} \right) erg \label{Erot} $. The consequent fraction of
energy lost after the flare must be severely bounded :
$\frac{\Delta(E_{Rot})}{E_{Flare}} \leq 10^{-6}$. Finally a signal
of secondary muons at PeV energies, induced by high energy
neutrinos from the SGR, might be detected  in Amanda and also
(because of its better orientation) in Baikal. To conclude, we
imagine that if the precessing jet model gives a correct
interpretation of the properties of SGRs, SGR 1806-20 will (and
indeed it has been) active all during 2005. Moreover, following
our prediction, the recent GRB060218 event showed long and short
scale quasi-periodic re-brightening, reflecting (as in GRB030329)
the inner multi-precessing jet. Part of this multi-bump signature
has been  and is still being  discovered in the new GRB events (
\cite{Fargion-GNC}) \footnote{A few very recent discovers on
peculiar GRBs and SGRs, after the Vulcano presentation in May
2005, have been introduced here to update and complete the
article. }.

\begin{figure}
\includegraphics[width=2.4in,height=2.4in]{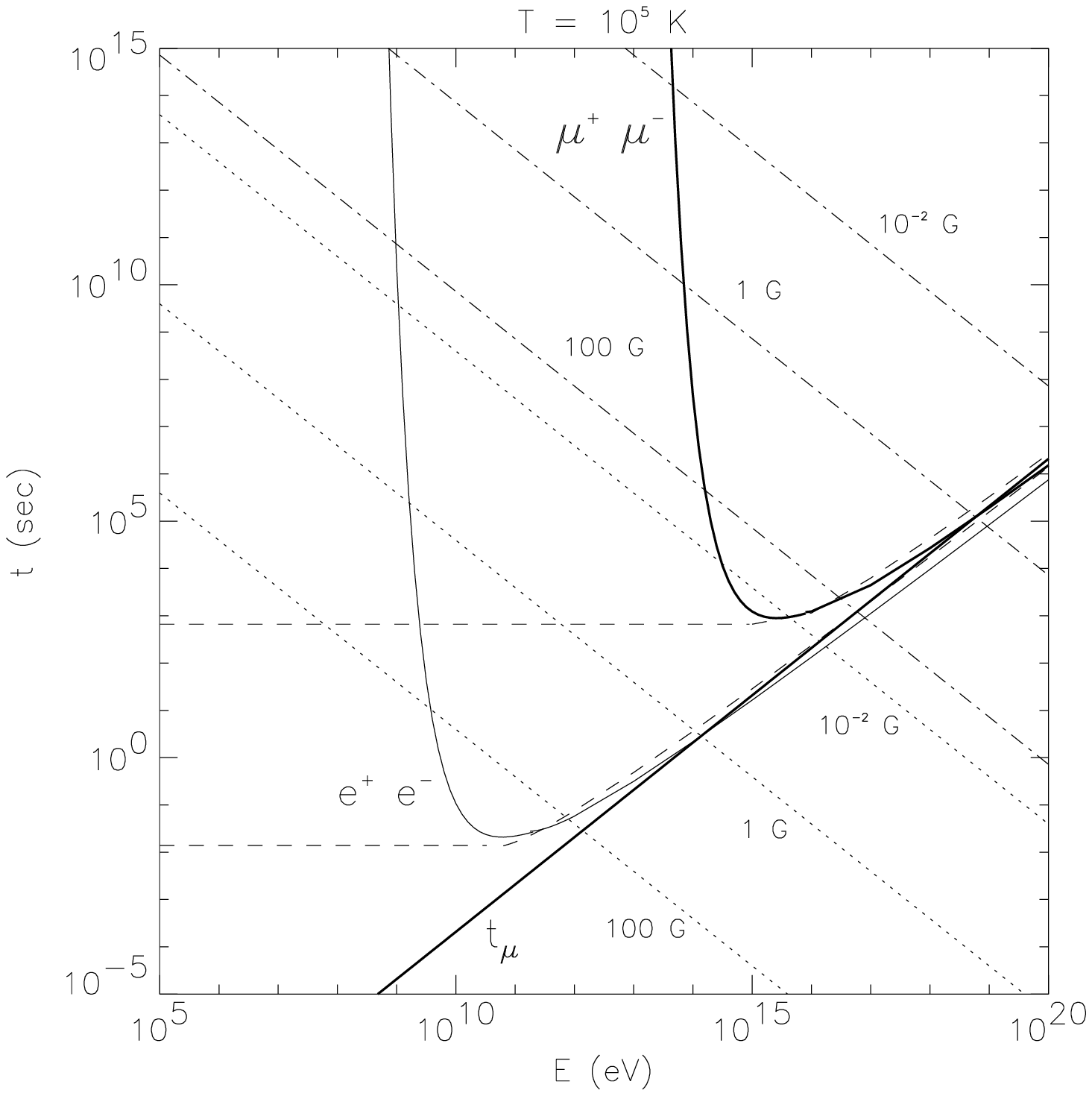}
\includegraphics[width=2.4in,height=2.4in]{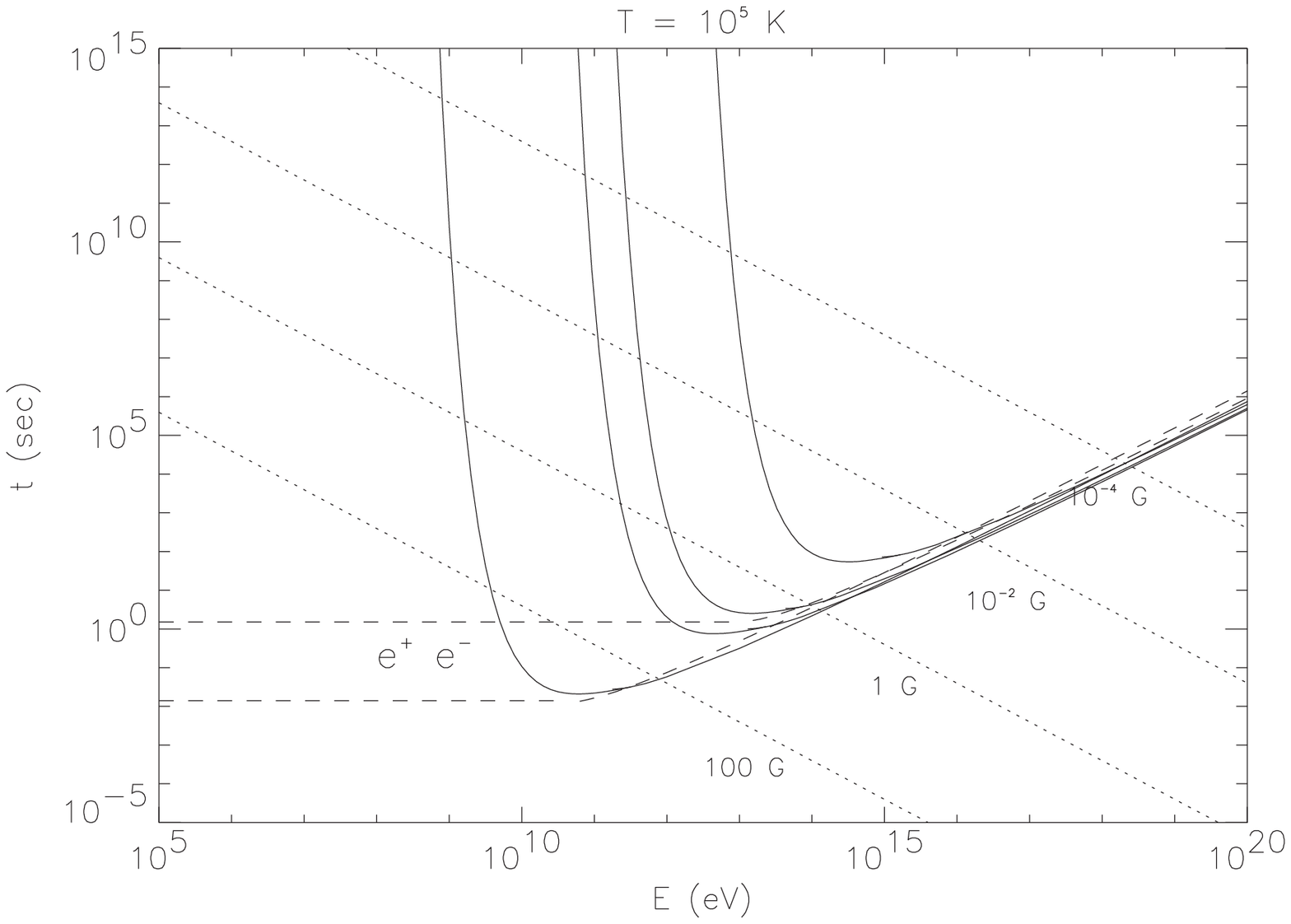}
 \caption{Left:The electron and muon interaction lengths. The {\em
dashed-dotted} and {\em dotted} lines correspond to the
synchrotron energy loss distance (for muons  and electrons
respectively) for different values of the magnetic field: 100 G, 1
G and $10^{-2}$ G. The {\em straight solid} line labelled
$t_{\mu}$ indicates the muon lifetime; the {\em dashed} lines
indicate the IC interaction lengths for muons and electrons.
Finally the two {\em solid} curves labelled $\mu^+ \mu^-$ and $e^+
e^-$ correspond to the attenuation length of high energy photons
producing lepton pairs (either $\mu^{\pm}$ or $e^{\pm}$) through
the interaction with the SN radiation field.  We have assumed that
the thermal photons emitted by the star in a pre-SN phase have a
black body distribution with a temperature $T \simeq 10^5$ K.
Assuming a radius $R \sim 10 \, R_{\odot}$, we are considering a
luminosity of $L_{SN} \simeq  2.5 \cdot 10^{41}$ erg  s$^{-1}$.
Around $10^{15} - 10^{16}$ eV muons decay before losing energy via
IC scattering with the stellar background or via synchrotron
radiation. Right :The Supernova opacity (interaction length) for
PeV electrons at different times. PeV muon jets may overcome it
and decay later in $\gamma$ showering electrons (see for details
Fargion, Grossi 2005); } \label{EEV-SGR}
\end{figure}

\begin{figure}
\includegraphics[width=2.5in]{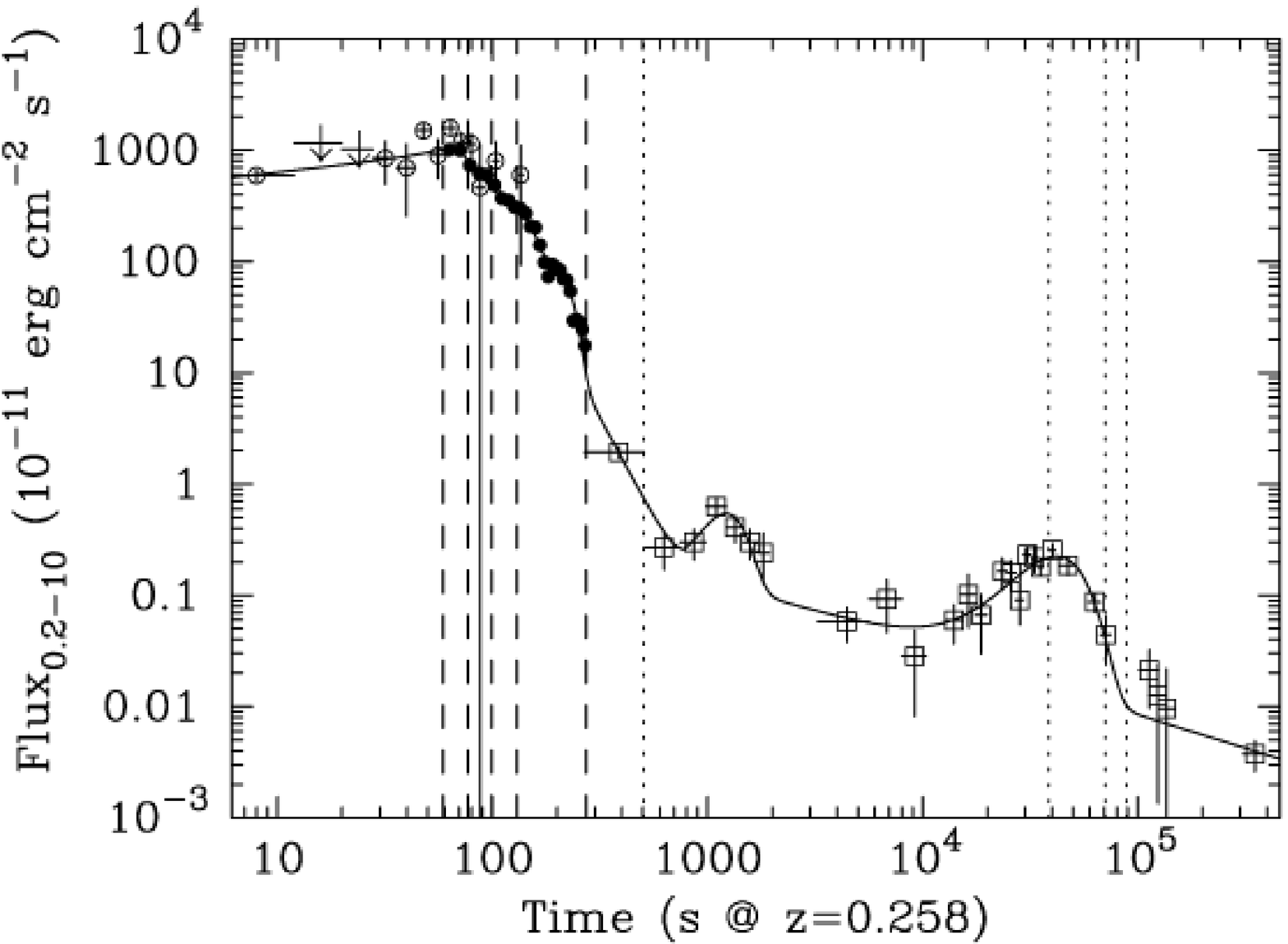}
\includegraphics[width=2.5in]{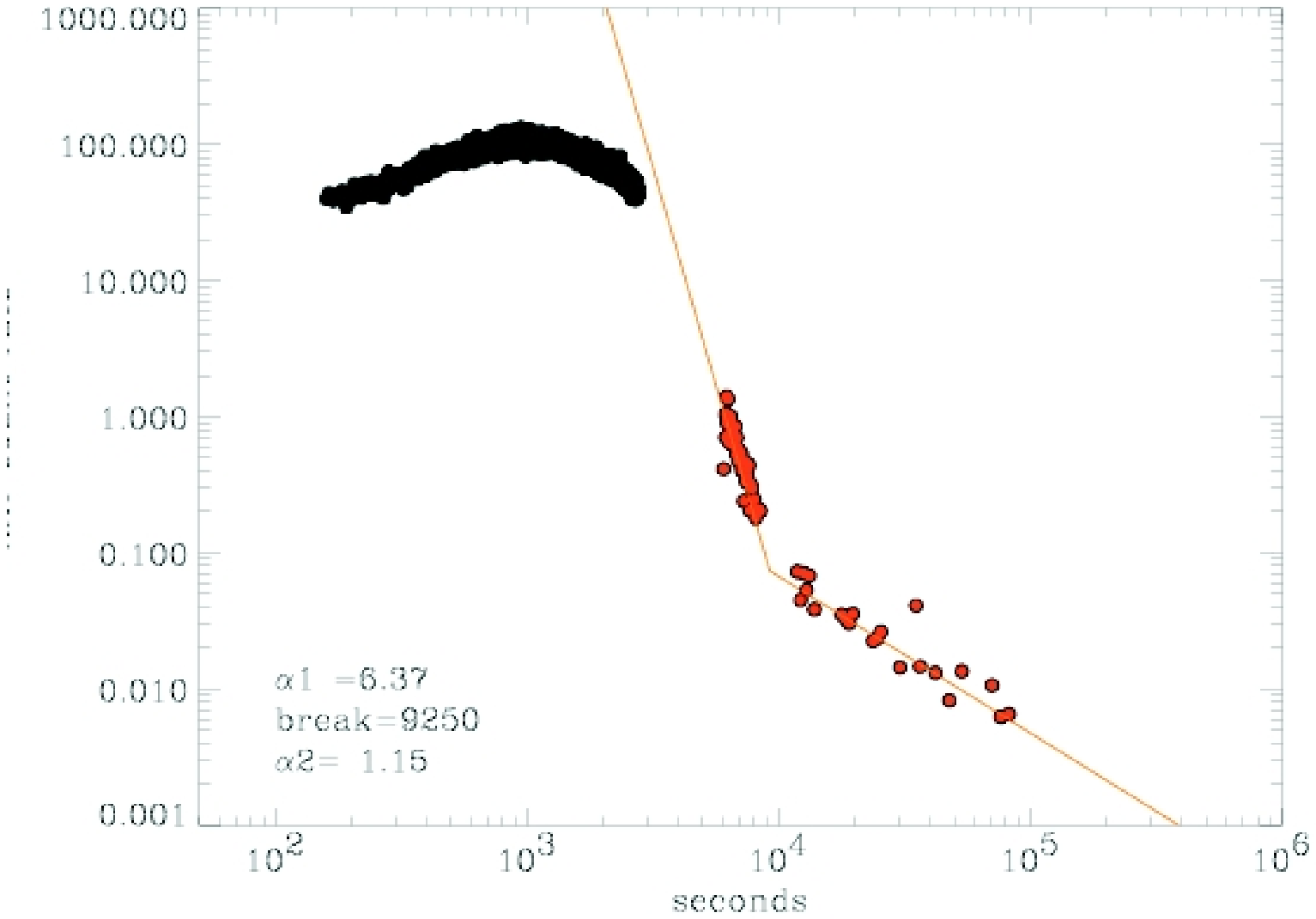}
 \caption{Left: The short GRB050724 and its long life X-ray afterglow   whose curve
 (\cite{Campana}) and whose multi-rebrightening is testing the persistent jet
 activity and  geometrical blazing views.
 Right: the very recent optical afterglows of GRB 060218 whose smooth longest X-ray flare and whose recent dramatic optical bumps
 are reflecting the underneath precessing jet activity
 (\cite{Moretti}; \cite{Fargion-GNC}).}
\end{figure}

\section{Conclusions: the Precessing Jet  answers}
 The thin precessing Jet while being extremely collimated (solid angle $\frac{\Omega}{\Delta\Omega}\simeq 10^{8}-10^{10}$ (Fargion,Salis
1995;Fargion 1999; Fargion,Grossi 2005); ) may blaze at different
 angles within a wide energy range (inverse of $\frac{\Omega}{\Delta\Omega}\simeq 10^{8}-10^{10}$ ). The emission at different
wavelengths is more  intense and harder in the inner part of the
jet.
  The Jet cone spreads while it precesses, leading to a blazing  variability
    mostly dominated by a tiny angle bending. The jet inner structure is made by concentric gamma radiation
    cones  produced  by higher energy electron pairs with harder
    spectra.
     The outer shells are characterised by a lower energy
     radiation. 
     The concentric shape of this ideal jet is deformed while
     turning and rotating in angular precession, due to the
     different "inertia" of the electron Jet components: the
     inner hard core remains on-axis while the softer external
     cones, and rings, are coming later as a tail, in some analogy
     to the well known Doppler ring structure.
    The output power may
 exceed $\simeq 10^{8}$, explaining the extreme low observed
 output in GRB980425, an off-axis event, the long late off-axis gamma tail by  GRB060218,
  respect to the on-axis and more distant GRB990123 (as well as GRB050904). In this scenario the Amati relation is not a physical law,
  but  just a biassed  statistical  rule that selects the most distant
  events  at large redshift,
  as the ones at peak activity with the best collimation, while gamma
  detectors may capture signals from more and more off-axis sources in the local universe or in our galaxy. The last GRB060218 following the Amati correlation is
  very off-axis and would not be easily observed at large redshifts;
    there is un undetected huge rate (few each second) of under-luminous  (five or six
    orders
    of magnitude less energetic) off-axis
    GRB in the Universe mixed up with unnoticed  lower power extragalactic
    SGRs. Short GRB (as GRB050709) are just the late
    stage between GRB toward SGR. XRF are off-axis events.
    We expect that the bumpy GRB060218 will  rebright possibly in
    all wavelengths (the longer the wavelength, the wider the cone) 
    as well as in delayed radio bumps.
    SGR1806-20 is an active jet whose beaming is rarely on-axis.
    Its precessing jet cone is sometimes blazing the Earth.
          In our galaxy there are hundreds of such precessing jets
    almost mixed up between AXRPs and SGRs; most of them have a jet cone  not
    pointing to us , like SS433 and other microquasars as Eta Carina jets feeding its twin  lobe.
    Therefore antiperiodic blazing are able to explain precursors
    and GRB or, viceversa a GRB and late X-afterglow rebrightening.
   Very recent GRB at cosmic edges are showing more bumps and
   variability as well as the very nearby GRB060218. Let us try to
   search for  jet traces in SGR 1806-20 and  in extragalactic nearby
   GRB060218: GRBs are not the most powerful explosions, but the most
   collimated ones.

\end{document}